\documentstyle[aps,floats,preprint,tighten]{revtex}
\input{psfig.tex}
\begin{document}
\draft
\preprint{\vbox{\hbox{FERMILAB--Pub--95/071-A}
\hbox{astro-ph/9504071}}}

\title{CBR Anisotropy and the Running of the Scalar
Spectral Index}

\author{Arthur Kosowsky}
\address{Harvard-Smithsonian Center for Astrophysics,
60 Garden Street, Cambridge, MA~~02138
\\and\\
Department of Physics, Lyman Laboratory, Harvard University,
Cambridge, MA~~02138}
\author{Michael S.~Turner}
\address{Departments of Physics and of Astronomy \& Astrophysics\\
Enrico Fermi Institute, The University of Chicago, Chicago, IL~~60637-1433
\\and\\
NASA/Fermilab Astrophysics Center\\
Fermi National Accelerator Laboratory, Batavia, IL~~60510-0500}
\date{April, 1995}
\maketitle

\begin{abstract}

Accurate ($\lesssim 1\% $) predictions for the anisotropy
of the Cosmic Background Radiation (CBR) are
essential for using future high-resolution ($\lesssim 1^\circ$) CBR
maps to test cosmological models.
In many inflationary models the variation (``running'') of the
spectral index of the spectrum
of density perturbations is a significant effect and leads to
changes of around 1\% to 10\% in the CBR power spectrum.  We propose
a general method for taking running into account which uses
the derivative of the spectral index ($dn/d\ln k$).
Conversely, high-resolution CBR maps may be able to determine
$dn/d\ln k$, giving important information about the inflationary potential.

\end{abstract}


The cosmic background radiation contains a wealth of information
about the spectrum of primeval density perturbations.  This is because
CBR anisotropy on a given angular scale
arises largely due to density perturbations on
a (comoving) length scale $L\simeq
(\theta /1^\circ )100h^{-1}\,{\rm Mpc}$.
Since the COBE detection of CBR anisotropy on angular scales of $10^\circ$
to $90^\circ$ \cite{dmr}, more than ten additional detections
on angular scales from about $0.5^\circ$ to $20^\circ$ have
been reported \cite{wss}.  In addition, plans are being made
for a satellite-borne experiment within the decade that
will map the CBR sky with an angular resolution of better than $1^\circ$
and an accuracy that is an order of magnitude better than current
measurements \cite{cobe2}.  Thus, in the near future CBR anisotropy
should be able probe inhomogeneity on length scales from about
$30h^{-1}\,{\rm Mpc}$ to $30,000h^{-1}\,{\rm Mpc}$.

A key to using CBR measurements
to reveal the underlying spectrum of density perturbations is the
accurate calculation of the expected anisotropy in a given model.
Much progress has been made in understanding and taking into account
all the relevant microphysics \cite{hu},
and several groups are now making a concerted effort
to calculate expected CBR anisotropies with
an accuracy of better than $1\%$ \cite{COMBA}.

Much of this effort is directed at inflation, as CBR anisotropy
has the potential to both test the inflation hypothesis
and reveal important information
about the underlying scalar-field potential \cite{recons}.  Inflationary
models predict approximately scale-invariant spectra of
density (scalar metric) perturbations \cite{scalar}
and gravity-wave (tensor metric) perturbations \cite{tensor}, and
both contribute to CBR anisotropy.
The following parameters have been identified as important
for accurately computing the expected anisotropy \cite{confusion}:
the power-law indices of the scalar and tensor spectra, $n_T\approx
0$ and $n\approx 1$; the overall amplitudes of the scalar
and tensor perturbations, often quantified by their contributions to the
variance of the quadrupole anisotropy, $Q_S$ and $Q_T$;
the Hubble parameter $h = H_0/100\,{\rm km\,s^{-1}}\,{\rm Mpc}^{-1}$;
the baryon density, which is constrained by primordial
nucleosynthesis to the interval $\Omega_Bh^2 \simeq
0.009 - 0.022$ \cite{cst}; and possible contribution of a cosmological
constant to the energy density of the Universe today
$\Omega_\Lambda$.  (In addition,
some have considered the possibility of a total energy density
less than the critical density predicted by almost all models
of inflation, nonstandard ionization histories for the Universe,
and variations in the nonbaryonic component of
the matter density, e.g., adding a small amount of hot dark matter.)

In this paper we emphasize that the spectral
indices $n$ and $n_T$ in general vary with scale and point out
that for many interesting models of inflation (chaotic, natural, and new)
the variation in scalar spectral index
leads to significant corrections (1\% to 10\%)
in the predicted CBR anisotropy.  Conversely, this means
that a high-resolution CBR map could be used to extract information about the
variation of $n$ with scale and thereby reveal additional information about
the inflationary potential.  We thus
make the case that the variation of the scalar spectral
index should be taken into account when calculating CBR anisotropy,
and suggest that it is most sensibly done by using $dn/d\ln k$.

CBR anisotropy on the sky is usually expanded in spherical harmonics,
\begin{equation}
\delta T (\Omega ) /T = \sum_{lm} a_{lm}Y_{lm}(\Omega ).
\end{equation}
Inflation makes predictions about the statistical properties
of the multipole moments; since
isotropy in the mean guarantees that $\langle a_{lm} \rangle =
0$ and the underlying perturbations in almost all inflationary
models are gaussian, the variance $C_l \equiv \langle |a_{lm}|^2
\rangle$ serves to specify all statistical properties.  (Here and
throughout brackets refer to the average over an ensemble of observers.)
Measurements of the CBR temperature on the sky can be used to
estimate the statistical properties of the underlying density
perturbations.  In particular, the $C_l$'s can be estimated.
Because the sky is but a finite sample, a fundamental limit
to the accuracy of the estimate (referred to as cosmic variance) is given by
\begin{equation}
\langle (C_l -C_l^{\rm est})^2\rangle = {2 C_l^2 \over 2l+1}.
\end{equation}
Other major (and presently dominant) sources of uncertainty include
receiver noise, various instrumental systematic errors,
foreground sources (our own galaxy, radio sources, etc.),
limited sky coverage,
and finite resolution (a map with angular resolution $\theta$ is only
sensitive to multipoles with $l\lesssim 200^\circ /\theta$).

High-resolution maps of the CBR probably offer the best means of
studying the scalar and tensor metric perturbations predicted
by inflation \cite{knoxmst,knox}.  Such maps may also
provide valuable information about the Hubble constant $H_0$,
the cosmological constant $\Lambda$, the baryon density $\Omega_B$,
and the total density of the Universe $\Omega$; however, other
measurements will complement the determination of these parameters.
If the four parameters describing the scalar and tensor perturbations
are measured to some level of accuracy, properties of the underlying
inflationary potential $V(\phi)$ can be determined \cite{recons1}:
\begin{eqnarray}
V_{N} & = & 1.65 Q_T\, {m_{\rm Pl}}^4  , \\
V_{N}^\prime & = & \pm \sqrt{- 8\pi n_T}\,V_{N}/{m_{\rm Pl}}
= \pm \sqrt{8\pi r \over 7}\, V_{N}/{m_{\rm Pl}} , \\
V_{N}^{\prime\prime} & = & 4\pi [(n-1) - 3n_T]\,V_{N}/{m_{\rm Pl}}^2
= 4\pi \left[ (n-1) + {3\over 7} r \right]\,V_{N} /{m_{\rm Pl}}^2 ,
\label{lowestorder}
\end{eqnarray}
where $r\equiv Q_T/Q_S$, a prime indicates derivative
with respect to $\phi$, and the sign of $V^\prime$ is
indeterminate.  In addition, a consistency relation $n_T = -r/7$
must be satisfied, and the factors of $1\over 7$ arise
from using it \cite{lambdanote}.
Subscript $N$ indicates that the potential is to be evaluated
at the value of $\phi$ where the scale corresponding to the present
Hubble scale ($k_N = H_0 $) crossed outside
the horizon during inflation.  This generally occurs around
$N\simeq 50$ e-foldings before the end of inflation, though the precise
expression depends upon the model of inflation, the
reheat temperature, and any entropy production after inflation.  (Only the
expression for $V_N$ depends upon the definition of $N$;
the other two always apply.)  The expression for $N$ can be
written as
\begin{equation}
N \simeq 54 + {1\over 6}\ln (-n_T) + {1\over 3}\ln
        (T_{\rm RH}/10^6\,{\rm GeV}) -{1\over 3}\ln \gamma -\ln h,
\end{equation}
where $T_{\rm RH}$ is the reheat temperature, $\gamma$ is the
ratio of the entropy per comoving volume today to that after reheating
which quantifies any post-inflation entropy production, and the
perturbation spectrum has been normalized to COBE.  (In calculating
$N$ it has been assumed that inflation is followed immediately
by a matter-dominated epoch associated with coherent oscillations of
the inflaton field and then by reheating.)

The above expressions were derived in a systematic approximation
scheme that relates the derivatives of an arbitrary
smooth inflationary potential to CBR observables \cite{recons1,reconnote}.
The expansion parameter is the deviation from
scale invariance, and formally involves all the derivatives
of the potential, ${m_{\rm Pl}}^{n}V_{N}^{(n)}/V_{N}$
(constant $V$ corresponds to the scale-invariant limit; see Ref.~\cite{lt}
for a discussion of this scheme).  For most potentials the deviation
from scale invariance of the scalar and tensor spectra,
quantified by $(n-1)$ and $n_T$, serve as the expansion parameters.
The above expressions are given to lowest order in $(n-1)$ and $n_T$;
the next-order corrections are given in Ref.~\cite{lt}.

{\it The crucial point for the present discussion is that the spectra
of scalar and tensor perturbations are only exactly power laws
for an exponential potential.}  In general, they vary with
scale, though $dn/d\ln k$ and $dn_T/d\ln k$ are second order in
the deviation from scale invariance, i.e., involve terms that
are ${\cal O}[(n-1)^2, n_T^2,
(n-1)n_T]$.  Since the present data indicate that scalar
perturbations do not differ from scale invariance by a
large amount, $n-1 = 0.10 \pm 0.32$ \cite{index},
the variation of the spectral indices is expected to be small.  Further,
indications are that the tensor perturbations are subdominant and
in any case only contribute significantly to multipoles $l=2$ to $50$.
However, we shall show that the variation of scalar spectral
index is important, given the desired precision for the
theoretical predictions of the multipoles.

The power spectrum for the scalar perturbations is given by
\begin{equation}
P(k) \equiv Ak^{n(k)} \propto \left( k\over k_N \right)^{n_N\, +\, \ln(k/k_N)
(dn/d\ln k)\, +\, \cdots }.
\label{kscaling}
\end{equation}
The contribution to the $l$th multipole comes from wavenumbers $k$ centered
around $l/\tau_0$, where $\tau_0 \simeq 2/H_0$ is the
distance to the last scattering surface.  Recalling that
the characteristic scale $k_N$ was chosen to correspond to
the current horizon size, this implies an approximate
scaling relation for the $C_l$'s which relates them to
a spectrum with constant spectral index:
\begin{equation}
C_l [n(k)] \simeq
\left( {l\over 2} \right)^{\ln (l/2)dn/d\ln k} \,C_l [n(k)=n_N].
\label{lscaling}
\end{equation}
If $| dn/d\ln k |\gtrsim 3\times 10^{-4}$, the effect of ignoring
the ``running'' of the spectral index over the range $l=2-1000$
is greater than one percent, which is significant compared to the
accuracy goal for CBR anisotropy \cite{COMBA}.  We now
show that values this large are expected in interesting inflationary models.

In general, the derivatives of the scalar and tensor spectral
indices are related to the inflationary potential and its
derivatives.  The lowest-order expression for {\it all} the
derivatives of $n$ and $n_T$ can be obtained by simply differentiating
the lowest-order expressions,
\begin{eqnarray}
n_T & = &  -{1\over 8\pi} \left( {m_{\rm Pl} V^\prime \over V} \right)^2
\Biggm|_{\phi=\phi_N} \\
n - 1 & = & n_T +{m_{\rm Pl}\over 4\pi}{d\over d\phi}\left( {m_{\rm Pl}
V^\prime\over V} \right)\Biggm|_{\phi=\phi_N},
\end{eqnarray}
using the fact that to lowest order
$${d\over d\ln k} = -{1\over 8\pi}\left({m_{\rm Pl}^2 V^\prime_N \over V_N}
{d\over d\phi}\right)\Biggm|_{\phi=\phi_N}.$$
If $n$ and $n_T$ are expressed as a function of $N$,
one can use the fact that $d/d\ln k = -d/dN$ to obtain the desired
derivatives even more easily.

It is thus a simple matter to obtain the first derivatives of $n$ and $n_T$:
\begin{eqnarray}
{dn\over d\ln k} & = &
-{1\over 32\pi^2}\left({{m_{\rm Pl}}^3V^{\prime\prime\prime}\over V}\right)
\left({{m_{\rm Pl}} V^\prime\over V}\right)
+ {1\over 8\pi^2}
\left({{m_{\rm Pl}}^2V^{\prime\prime}\over V}\right)
\left({{m_{\rm Pl}} V^\prime\over V}\right)^2
- {3\over 32\pi^2}\left({m_{\rm Pl}} {V^\prime\over V}\right)^4 \\
&\phantom{=}& \qquad\qquad {dn_T\over d\ln k} =
{1\over 32\pi^2}\left({{m_{\rm Pl}}^2V^{\prime\prime}\over V}
\right) \left( {{m_{\rm Pl}} V^\prime\over V}\right)^2  - {1\over 32\pi^2}
\left({{m_{\rm Pl}} V^\prime\over V}\right)^4.
\end{eqnarray}
Equivalent expressions can
be obtained by using the previous equations relating $r$ and
$n-1$ to the potential and its first two derivatives:
\begin{eqnarray}
{dn\over d\ln k} & = & \mp {1\over 16\pi^2} \sqrt{2\pi \over 7} \,
\left( {{m_{\rm Pl}}^3 V^{\prime\prime\prime}_{N}\over
V_{N}}\right) \, \sqrt{r} + {4\over 7} (n_{N} -1)r + {6\over 49} r^2, \\
{dn_T\over d\ln k} & = & -n_T[ (n-1) - n_T] = {r\over 7} \left[(n-1)
- {1\over 7}r \right] ,
\end{eqnarray}
where the upper sign applies if $V^\prime_{N} >0$ and
the lower if $V^\prime_{N}<0$, and the factors of $1\over 7$
arise from using the consistency relation $n_T =-r/7$.
{}From these expressions
we see that the size of both $dn/d\ln k$ and $dn_T/d\ln k$ is controlled by
the ratio of tensor to scalar perturbations, and further,
that the size of $dn_T/d\ln k$ depends upon the difference
between $n-1$ and $n_T$, which in many models is small.

We now quantify expectations in several popular models of inflation.
As noted earlier, for an exponential potential $dn/d\ln k =
dn_T/d\ln k \equiv 0$.
For inflation models that are based upon Coleman-Weinberg like
potentials, $V(\phi ) = B\sigma^4/2 + B\phi^4 [\ln (\phi^2/\sigma^2) - 1/2]$,
\begin{eqnarray}
dn/d\ln k & \simeq & -1.2\times 10^{-3}(50 /N)^2, \\
d^m n/d\ln k^m & \simeq & -3m!/N^{m+1}.
\end{eqnarray}
Chaotic-inflation models are
usually based upon potentials of the form $V(\phi ) = a \phi^b$
($a$ is a constant and $b=2,4, \cdots$ is an even integer) and
\begin{eqnarray}
dn /d\ln k & = & -4\times 10^{-4}\, (b/2+1)(50/N)^2, \\
d^m n/d\ln k^m & = & -m!\,(b/2+1)/N^{m+1}.
\end{eqnarray}
For the interesting cases of $b=2$ and $4$,
$dn/d\ln k = -0.8\times 10^{-3}$
($b=2$, $N=50$) and $-1.2\times 10^{-3}$ ($b=4$, $N=50$).
Finally, for the ``natural'' inflation model, where $V(\phi ) =
\Lambda^4 [1 + \cos (\phi /f) ]$, the following approximate
expression applies for $f\lesssim {m_{\rm Pl}}$ (which is the regime where
the deviation from scale invariance is significant and
$1-n \simeq {m_{\rm Pl}}^2/8\pi f^2$):
\begin{equation}
dn/d\ln k = -{\pi^2 \over 4}(n-1)^2 \exp [N(n-1)].
\end{equation}
Varying $(1-n)$ from $0.04$ to $0.3$ and $N$ from $40$ to $50$,
$dn/d\ln k$ varies from about $-10^{-7}$ to almost $-10^{-3}$.
Even though $1-n$ can be large in these models, $r$ is very
small when it is, and $dn /d\ln k$ never approaches $(1-n)^2$.

It is not a complete surprise that $dn/d\ln k$ is similar in
all these models.  On naive grounds one might expect that
$(n-1) \propto 1/N^m$, so that
$dn/d\ln k = -dn/dN = m (n-1)/N$.  This is true for
new and chaotic inflation where $m=1$.  (As usual, the situation
with ``natural'' inflation is more complicated.)

Are there models where running is more important?
For an ad hoc potential the answer is yes; consider $V(\phi )=
V_0 \exp (-a\phi^b)$.  Here
$$(n-1) = - {8\pi \over (2-b)^\alpha}\left[{ab\over 8\pi}\right]^{2-\alpha}
\,{1\over N^\alpha}  + {\alpha \over N}, $$
where $\alpha = 2(1-b)/(2-b)$.  For $b\not= 1$, $n-1$ can be large
and $dn/d\ln k \simeq \alpha(n-1)/N$.

As mentioned earlier, the running of the tensor spectral
index is expected to be less important because the tensor
perturbations are likely to be subdominant and only contribute
significantly for $l\lesssim 50$; in addition, $dn_T/d\ln k$
is smaller (being proportional to the difference
between $n-1$ and $n_T$ which is often small).  For the
potentials discussed above $dn_T/d\ln k =0$ (exponential), $-2\times
10^{-7}(\sigma /{m_{\rm Pl}})^4(50/N)^4$ (new),
$-2.0\times 10^{-4}b(50/N)^2$ (chaotic),
and $-{\pi^2/4}(n-1)^2 \exp [N(n-1)]$ (``natural'').

Figure 1 displays the CBR angular power spectrum for
$b=6$ chaotic inflation (where $dn/d\ln k= -0.0020$), calculated
without and with the running of the scalar spectral
index.  The correction due to the running of scalar index
is significant (about 10\%) and potentially measurable.
The results shown have been calculated using the
power spectrum in Eq.~(\ref{kscaling}).  We have also
calculated the $C_l$'s using the approximation in
Eq.~(\ref{lscaling}), and the maximum error in any $C_l$ is less than 0.6\%.
Thus, for applications requiring ${\cal O}(1\%)$ accuracy, it
should be sufficient to calculate a model with a fixed $n$ and then
scale the results according to Eq.~(\ref{lscaling}) to obtain
$C_l$'s for $dn/d\ln k \not= 0$.
We also note that the correct $k$-space power spectrum is
simple to include in any Boltzmann code.

In summary, expectations for $|dn /d\ln k|$ in popular
inflationary models range from $-2\times
10^{-3}$ to around $-4\times 10^{-4}$.  Of course, the
value of $dn/d\ln k$ in ``the model of inflation'' could
be larger or smaller.  At the high end of this range, neglecting the
running of scalar spectral index leads to errors of 10\%,
more than an order of magnitude larger than the accuracy
desired \cite{whitecomment}.  The running of the
scalar spectral index can be into account easily, accurately, and with
generality by using $dn/d\ln k$. Based upon the models we have looked
at one could adopt $dn/d\ln k \simeq (n-1)/N$ as a default estimate.

If the running of the
scalar spectral index is large enough to detect,
the third derivative of the scalar potential can be measured \cite{lt}:
\begin{equation}
V_{N}^{\prime\prime\prime} /{m_{\rm Pl}} = \pm 39 \sqrt{r}\left[
-7 (dn/d\ln k )/r + 0.9 r + 4 (n-1) \right]\,Q_T .
\end{equation}
The feasibility of determining $dn/d\ln k$ from a
high-resolution map of the CBR sky is currently under study \cite{jkk}.

Two final points.  First, what about the next-order corrections?
They involve ${\cal O}[(n-1)^3, n_T^3, \cdots]$ terms:
corrections to $dn/d\ln k$, $n$, $Q_T$, $Q_S$
and the $d^2n/d\ln k^2$ term in the expansion for $n$.   Provided that
the deviation from scale invariance is not too large,
they should be small (less than about 1\%)
because they are suppressed by an additional
factor of ${\cal O}[(n-1), n_T]$; e.g., $dn^2/d\ln k^2 = -5\times
10^{-5}(50/N)^3$ for new inflation, which leads to a correction at
$l=1000$ of about 0.5\%.  If the $d^2n/d\ln k^2$ should be
larger, its size might be turned to good purpose; because of
the qualitative difference between it and the $d/d\ln k$ term
it might possibly be measured, revealing information about the
fourth derivative of $V(\phi )$.

Last, but perhaps not least, the running
of the scalar spectral index is also of some relevance when
extrapolating a COBE-normalized spectrum to astrophysical
scales; e.g., the correction to $\sigma_8$ is about
$-3\%$ for $dn/d\ln k=-10^{-3}$ \cite{kofman}.

\acknowledgments

We thank Scott Dodelson, Paul Steinhardt, and Martin White
for thoughtful comments and Ed Copeland, Andrew Liddle, and Jim Lidsey
for pointing out errors in an earlier version of this manuscript.
Marc Kamionkowski and Gerard Jungman developed the code used to
calculate the power spectra in Fig.~1.
This work was supported in part by the DOE (at Chicago and
Fermilab) and the NASA (at Fermilab through grant NAG 5-2788).
AK is supported by the Harvard Society of Fellows.

\vfil\eject

\begin{figure}[t]
\centerline{\psfig{figure=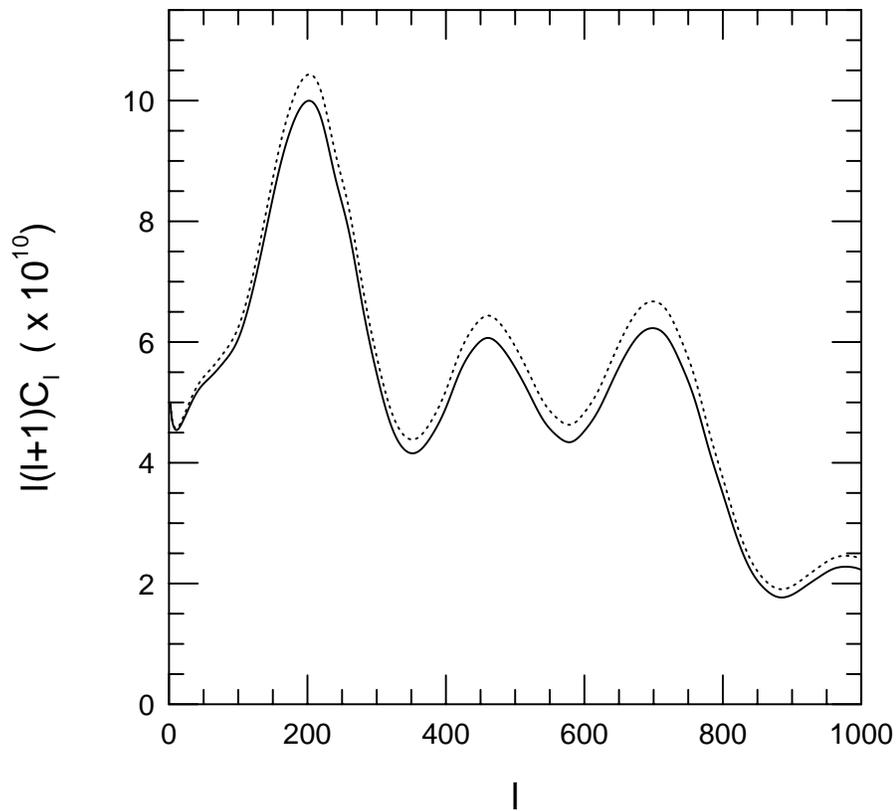,height=8in}}
\caption{ Predicted angular power spectra
for $b=6$ chaotic inflation with (solid) and without (broken) the running
of the scalar spectral index ($n=0.92$, $h=0.7$, $\Omega_B = 0.025$, and
$dn/d\ln k = -0.002$). }
\end{figure}


\begin{references}
\bibitem{cobe2}  Presently three groups in the US (FIRE, MAP, and
PSI) are preparing proposals for an Explorer-class NASA
satellite to carry out high-resolution mapping of the CBR;
the COBRAS/SAMBA collaboration in Europe is preparing a similar
proposal for an ESA mission.
\bibitem{lambdanote} For a nonzero cosmological constant this
relationship is modified slightly; see Ref.~\cite{knox}.
\bibitem{reconnote}  The virtue of the expansion scheme used here is
its generality.  For specific potentials other approximations
can be useful.  For example, for new inflation and $\sigma\ll
m_{\rm PL}$, $n_T \approx 0$ and $|\delta_k|^2 \propto
k [N - \ln (k/k_N)]^3$.
\end{references}
\end{document}